\documentclass[11pt,a4paper]{amsart}
\usepackage{amsmath}
\usepackage{amssymb}
\usepackage{amsfonts}
\usepackage{hyperref}

\usepackage{epsfig}
\usepackage{epsf}

\usepackage{graphicx}
\usepackage{graphics}

\def\ii{{\,{\rm i}\,}}
\def\={\ =\ }
\def\dd{{\rm d}}

\newcommand{\Tr}[1]{\:{\rm Tr}\,#1}
\def\e{{\,\rm e}\,}
\newcommand{\mbf}[1]{{\boldsymbol {#1} }}
\newcommand{\IZ}{\mathbb{Z}}
\newcommand{\IC}{\mathbb{C}}

\newcommand{\IN}{\mathbb{N}}
\newcommand{\IR}{\mathbb{R}}

\def\appendix#1{\addtocounter{section}{1}\setcounter{equation}{0}
\renewcommand{\thesection}{\Alph{section}}
\section*{Appendix \thesection. #1}
\protect\indent \parbox[t]{11.715cm}
}

\newcommand{\beq}{\begin{eqnarray}}
\newcommand{\eeq}{\end{eqnarray}}

\newcommand{\Hcal}{\mathcal{H}}

\newcommand{\Qsf}{{\sf Q}}

\theoremstyle{plain}

\numberwithin{equation}{section}

\begin{document}
\title[Two-dimensional Yang-Mills theory and the six-vertex model]{Two-dimensional
Yang-Mills theory, \\ Painlev\'e equations and the six-vertex model}
\date{December 2011 \hfill\ HWM--11--3 \ , \ EMPG--11--04 \ }
\author{Richard J. Szabo}
\address{\flushleft Department of Mathematics\\ Heriot-Watt University\\ Colin
Maclaurin Building, Riccarton, Edinburgh EH14 4AS, UK\\ and Maxwell Institute
for Mathematical Sciences, Edinburgh, UK}
\email{R.J.Szabo@ma.hw.ac.uk}
\urladdr{}
\thanks{}
\author{Miguel Tierz}
\address{\flushleft Einstein Institute of Mathematics\\ The Hebrew University
of Jerusalem\\ Givat Ram, 91904 Jerusalem, Israel}
\email{tierz@math.huji.ac.il}
\address{\flushleft Grupo de F\'isica Matem\'atica\\ Complexo Interdisciplinar da Universidade de Lisboa 
\\ Av. Prof. Gama Pinto, 2, PT-1649-003 Lisboa, Portugal.}
\email{tierz@cii.fc.ul.pt}
\urladdr{}
\curraddr{ }

\begin{abstract}
We show that the chiral partition function of two-dimensional Yang-Mills theory on the
sphere can be mapped to the partition function of the homogeneous six-vertex model with domain wall boundary conditions in the ferroelectric
phase. A discrete matrix model description in both cases is given by the Meixner
ensemble, leading to a representation in terms of a stochastic
growth model. We show that the partition function is a particular case
of the $z$-measure on the set of Young diagrams, yielding a unitary matrix
model for chiral Yang-Mills theory on $S^{2}$ and the
identification of the partition function as a tau-function of the Painlev%
\'{e}~V equation. We describe the role played by generalized non-chiral Yang-Mills theory on $S^2$ in relating the Meixner matrix model to the Toda chain hierarchy encompassing the integrability of the six-vertex model. We also argue that the thermodynamic behaviour of the six-vertex model in the disordered and antiferroelectric phases are captured by particular $q$-deformations of two-dimensional Yang-Mills theory on the sphere.
\end{abstract}

\maketitle

\section{Introduction and summary of results}

\subsection{Gauge theories and statistical mechanics}

The relationships between gauge theories and exactly solvable models of statistical
mechanics has been a subject of intense activity for many years, in diverse dimensions and contexts. Such connections have the potential to unveil the dynamical reasons as to why only particular selected classes of systems are integrable.
In recent years this study has also shed light on particular tractable sectors of gauge theories. The
most distinctive example is the study of the planar limit of $\mathcal{N} = 4$ maximally
supersymmetric Yang-Mills theory in terms of integrable spin chains~\cite%
{Minahan:2002ve,Beisert:2003tq} (see~\cite{Beisert:2010jr} for a recent
review). In the context of the AdS/CFT correspondence, the integrability structures on both the string theory and gauge theory sides have been identified and matched. In this paper we present some new connections between certain integrable lattice models of statistical mechanics and two-dimensional Yang-Mills theory; this theory has a long history as an exactly solvable quantum gauge theory with deep connections to one-dimensional integrable systems of Calogero-type, string theory, topological field theory, and the geometry of moduli spaces.

\subsection{Two-dimensional Yang-Mills theory}

In this paper we study quantum Yang-Mills theory with gauge group $SU(N)$
on an oriented closed
Riemann surface $\Sigma$ of genus $h$ and unit area form $\dd\mu$~\cite{cordesmoore}. The action is%
\begin{equation}
S_{\rm YM}=-\frac{1}{4g_s}\, \int_{\Sigma}\,
\dd\mu~\mathrm{Tr}\, F^{2} \ ,
\label{continuum}
\end{equation}%
where $g_s$ plays the role of the coupling constant, $F$ is the field strength of a matrix gauge connection, and $\Tr$
is the trace in the fundamental representation of $SU(N)$.
It was Migdal's idea to utilize a lattice regularization of the gauge theory,
which relies on a triangulation of the two-dimensional manifold
$\Sigma$ with group matrices situated along the
edges~\cite{Migdal75}. The path integral is then approximated by
the finite-dimensional unitary matrix integral
\begin{equation}
\mathcal{Z}_{\rm M}=\int\, \prod_{{\rm edges}~\ell}\, \dd
U_{\ell}~ \prod\limits_{\mathrm{plaquettes}~P}\, Z_{P}\left[
U_{P}\right] \ ,
\label{latticeint}\end{equation}%
where $\dd U_\ell$ denotes Haar measure on $SU(N)$ and the holonomy
$U_{P}=\prod\nolimits_{\ell \in P}\, U_{\ell}$ is the ordered product of group matrices
along the links of a given plaquette. The local factor $Z_{P}\left[ U_{P}\right] $ is a
suitable gauge invariant lattice weight that converges in the continuum limit to the
Boltzmann weight for the Yang-Mills action (\ref{continuum}).

There
are two common choices for the lattice weight $Z_{P}\left[ U_{P}\right] $, involving the
Wilson action and the heat kernel action. The latter action has many interesting
features~\cite{Menotti:1981ry} and is the usual choice in two-dimensional Yang-Mills
theory~\cite{cordesmoore}. It leads to the well-known group theory expansion of the
partition function~\cite{Migdal75,Rusakov}%
\begin{equation}
\mathcal{Z}_{\rm M}=\sum_{\lambda }\, \left( \dim \lambda \right)
^{2-2h} \, \exp
\big( -g_{s}\,C_{2}(\lambda ) \big) \ ,  \label{HK}
\end{equation}%
where the sum runs through all isomorphism classes $\lambda $ of irreducible representations of the
$SU(N)$ gauge group, $\dim\lambda$ is the dimension of the
representation $\lambda $, and $C_{2}(\lambda )$ is the quadratic Casimir
invariant of $\lambda$.

In the lattice approximation one generally
expects that in the limit of a very fine triangulation we get
a theory that converges to the continuum gauge theory. In the present case, this approach is in fact much more powerful due to an
invariance property of the partition function (\ref{HK}) under subdivision of
plaquettes of the lattice, which is specific to two dimensions.
This property implies that the lattice computation is independent of
the chosen triangulation of $\Sigma$, and hence is exact on an arbitrarily
large lattice~\cite{Migdal75,Witten:1991we}. This opens up the
possibility that there may exist an integrable lattice model of two-dimensional statistical
mechanics that exactly describes, or is equivalent to, two-dimensional
Yang-Mills theory in the continuum. In this paper we
show that this is indeed the case
for the chiral sector of the gauge theory, which is defined as follows.

The sum in (\ref{HK}) runs through
all irreducible representations of $SU(N)$, but one can also restrict the sum
to a subclass of representations. In the large $N$ limit, any
representation $\lambda$ of $SU(N)$ can be expressed
uniquely~\cite{Gross:1993hu} in terms of coupled representations
$\lambda=\overline{\lambda_+}\,\underline{\otimes}\, \lambda_-$,
defined to be the largest irreducible representation in the
decomposition of the tensor product
$\overline{\lambda_+}\otimes\lambda_-$, such that the Young tableau
for $\lambda$ is given by joining a chiral tableau $\lambda_+$ to an
antichiral tableau $\lambda_-$. The number of boxes in the Young
tableaux corresponding to $\lambda_\pm$ are understood as being small
compared to $N$. Then the
Hilbert space $\mathcal{H}_{\rm YM}^{SU(N)}$ of class functions on
$SU(N)$ factorizes for
large~$N$ into the coupled tensor product of two sectors~\cite{Gross:1993hu,Gross:1993yt}%
\begin{equation}
\lim_{N\to\infty}\, \mathcal{H}^{SU(N)}_{\rm YM}= \mathcal{H}_{+}\, \underline{\otimes}\, \mathcal{H}%
_{-} \ . \label{fact}
\end{equation}%
The chiral Hilbert space $\mathcal{H}_+$ consists of states corresponding to ``small''
representations $\lambda_+$ of $SU(\infty)$ in which the number of
Young tableau boxes is an arbitrary but finite non-negative integer,
while the antichiral Hilbert space $\mathcal{H}_-$ consists of states
corresponding to conjugates of small representations.

The chiral partition function $Z_{\rm YM}^+(\Sigma,SU(N))$ on an arbitrary
surface $\Sigma$ is defined by keeping only the states of $\Hcal_+$ in
the large $N$ Hilbert space (\ref{fact}).
This definition also makes sense in the $q$-deformation of the gauge theory~\cite{Aganagic:2004js,Caporaso:2005fp}
that we additionally consider, whose heat kernel expansion is given by
substituting the dimensions $\dim\lambda$ of $SU(N)$ representations
with their quantum dimensions $\dim_q\lambda$ in (\ref{HK}). In this case the chiral/antichiral factorization is important in
the context of the OSV conjecture in topological string theory~\cite{Ooguri:2004zv}.

\subsection{Six-vertex model with domain wall boundary conditions}

In~\cite{Witten:1991we} Witten indicated a relationship
between two-dimensional Yang-Mills theory and IRF models,
in the presence of Wilson loops and in a rather generic way. He showed that
the lattice gauge theory description of Wilson line correlators could be expressed as a lattice statistical
mechanics formula similar to that of an IRF model. In this paper we
explore a different connection with integrable lattice models which are dual to IRF
models, i.e. six-vertex models. We show
that the partition function $Z_{\rm YM}^{+}\left({S}%
^{2}\,,\,SU(N)\right) $ for the
chiral sector of Yang-Mills theory on the two-sphere $\Sigma= S^{2}$ can be mapped to that of the six-vertex model with domain wall boundary
conditions in its ferroelectric regime~\cite%
{Pz}. The results of this paper give gauge theory derivations of
these lattice models which elucidate further integrability properties on both
the gauge theory and statistical mechanics sides. The gauge theories
may also provide computationally useful means for exploring various
aspects and for understanding the origins of these exactly solvable
models. Previous correspondences between non-chiral
two-dimensional Yang-Mills theory and certain one-dimensional integrable
systems are found in~%
\cite{Gorsky:1993pe,Minahan:1993mv,Szabo:2010qv}.

The six-vertex model is a two-dimensional exactly solvable lattice statistical mechanics model, introduced by Lieb
and Sutherland~\cite{Lieb1,Lieb2,Suth}, in which local
states are associated with edges of an $N\times N$ square lattice and
local statistical
weights are assigned to its vertices. The states each take two values that are
usually denoted as arrows along the edge. Since the lattice is square,
there are in principle 16 possible arrow configurations around each vertex,
but most of them are chosen to have zero weight in such a way
that only six configurations are allowed with equal numbers of incoming and outgoing
arrows. The partition function of the six-vertex model is then%
\begin{equation}
Z_{N}=\sum_{\stackrel{\scriptstyle \text{arrow}}{\scriptstyle
    \text{configurations }\sigma }}~ \prod\limits_{i=1}^{6}\, w_{i}^{N_{i}\left( \sigma \right) } \ ,
\label{ZN6V}\end{equation}
where $w_i$, $i=1,\dots,6$ are weights associated to each possible
vertex state and $N_i(\sigma)$ is the number of vertices of type $i$
in the configuration $\sigma$.

The six-vertex model suffers from an intricate dependence on boundary
conditions, due to the constraints imposed by arrow conservation. In particular, the free
energy computed with domain wall boundary conditions is different from
that computed with periodic
boundary conditions, even in the infinite volume limit~\cite{Kor1,IzKor,IzKor2}.
With domain wall boundary conditions, all arrows on the left and right boundaries
are outgoing, while on the top and bottom boundaries all arrows are
incoming. In addition to demonstrating the dependence of thermodynamic
quantities on the boundary conditions~\cite{KorPz,Pz}, there is currently
much interest in this model due to its deep connections with several problems
in algebraic combinatorics. In particular, Kuperberg studied its partition
function, and used it to give a direct and transparent proof of the
alternating sign matrix
conjecture~\cite{Kuperberg,Bressoud}. There is also a direct connection with
the study of domino tilings~\cite{KorPz}; in particular, the free fermion
line of the six-vertex model with domain wall boundary conditions is
related to the domino tiling problem for the Aztec diamond~\cite{Aztec}.

Recall the exact self-similarity property of
two-dimensional lattice gauge theory~\cite{Migdal75,Witten:1991we}. We will show that the
precise macroscopic two-dimensional lattice model that can be mapped to $SU(N)$ chiral Yang-Mills theory on $S^{2}$ is the six-vertex model with domain wall boundary
conditions in the ferroelectric phase, in the limit introduced in~\cite{Pz}. In the six-vertex
model the rank of the gauge group $N$ corresponds to the size of the $N\times N$ square
lattice; the mapping
between coupling parameters is given in (\ref{gs2tgamma}) below. We will also argue that the six-vertex model description beyond this approximation constitutes an interesting special case of generalized two-dimensional Yang-Mills theory. This gives a statistical mechanics interpretation of the chiral
sector of Yang-Mills theory, and also an interpretation of the partition
function in terms of the normalization constant of a certain stochastic process.
Conversely, the six-vertex model in this ordered phase and
with these boundary conditions can be described as a topological gauge
theory on a single plaquette.

\subsection{Unitary matrix models and Painlev\'e transcendents}

The partition function of two-dimensional Yang-Mills theory based on
the heat kernel
action (\ref{HK}), i.e. as a sum over irreducible $SU(N)$ representations $\lambda$, can be rewritten in the linearized chiral case in terms of the normalization constant of the
Schur measure~\cite{Szabo:2010sd}. This is especially
relevant for the chiral sector, because in this case it is
straightforward to write the partition function as a Toeplitz or
Fredholm determinant~\cite{Szabo:2010sd}. This determinant expression leads to connections with
unitary matrix models and with integrable hierarchies.

The first physical model whose correlation functions were expressed as a Toeplitz
determinant was the two-dimensional Ising model, arguably the simplest of all integrable
systems. The seminal work~\cite{McCoy:1976cd,McCoy:1977er} established that
the two-point correlation functions of spin and disorder fields can be expressed
in terms of a solution to the Painlev\'{e}~III equation. This result was
extended and formalized by the Kyoto school in~\cite{Kyoto}. In particular, they connected the theory of isomonodromy
preserving deformations of linear differential equations with the $n$-point
correlation functions of the two-dimensional Ising model, and also related the reduced
density matrix of the impenetrable Bose gas model with the Painlev\'{e}~V
transcendent. Other models, possessing a free fermion region, also have
correlation functions that solve non-linear differential equations~\cite%
{Its,book}. In this approach, a Fredholm determinant representation of
the correlators is crucial.

There are a great number of models whose correlation
functions are governed by a Painlev\'e transcendent; see~\cite{TWPainleve} for
a recent review. As pointed out in~\cite{TWPainleve}, in general it is not
expected that Painlev\'{e} transcendents arise in correlation functions of
models that are exactly solvable but which are not free fermion
models, such as the
six-vertex and eight-vertex models, and the XXZ quantum spin chain. There seem to be
exceptions that include certain ferromagnetic models. The present work
goes partly in this direction, as it relates the Painlev\'{e}~V
transcendent to the six-vertex
model with domain wall boundary conditions in the ferroelectric phase,
and in particular away from the free fermion line of the model. In
contrast, two-dimensional Yang-Mills theory does have
fermion operator representations~\cite{Minahan:1993tp,cordesmoore}; this has some implications for the usual matrix model description of the six-vertex model with domain wall boundary conditions in the ferroelectric phase~\cite{Pz}.

Another important and natural appearance of Painlev\'{e} transcendents is
in the reinterpretation of
two-dimensional quantum gravity in terms of matrix models. This approach led, in
the work of Br\'{e}zin and Kazakov, Douglas and Shenker, and Gross and
Migdal~\cite{bk,gm,ds}, to exact solutions of two-dimensional gravity
coupled with matter fields. A relationship between the free energy in the
double-scaling limit of the multicritical matrix models and solutions
of the Painlev\'{e}~I equation was a crucial result. The role of the discrete Painlev\'e transcendents
and the tau-function of an isomonodromic deformation in two-dimensional quantum gravity was later studied in
further detail in~\cite{Moore:1990mg,Fokas:1990wb,Fokas:1991za}.

The double scaling limit of the full heat kernel expansion for Yang-Mills theory on the sphere is related to Painlev\'e~II~\cite{Gross:1994mr}. In this case the double scaling free energy $F_{\rm YM}(t)$ is given by $F_{\rm YM}^{\prime\prime}(t)=v(t)^2/4$, where $v(t)$ satisfies the Painlev\'e~II equation
$$
2v^{\prime\prime}-v^3+t\, v=0 \ .
$$
The same property is shared by the Gross-Witten model~\cite{Periwal:1990gf}; this is another combinatorial model of two-dimensional quantum Yang-Mills theory obtained as a one-plaquette model based on the Wilson action instead of the heat kernel action in (\ref{latticeint}). While Painlev\'e~I appears as the universality class of two-dimensional gravity, the Painlev\'e~II equation describes two-dimensional supergravity~\cite{Kms}. In the Gross-Witten model the relationship with Painlev\'e transcendents goes beyond the connection between the specific heat and Painlev\'e~II in the double scaling limit, and it involves the Painlev\'e~III and Painlev\'e~V equations as well~\cite{Hisakado:1996di,TW}. The appearance of one equation or the other depends on the matrix model quantity being considered and whether or not a double scaling limit is taken; we summarize these results in Appendix~A. This feature is particularly important in our situation, because by considering the works~\cite{Bor,ForrWitte} we can establish analogous results for chiral Yang-Mills theory on $S^2$ and also for the six-vertex model with domain wall boundary conditions.

Both the six-vertex model and the chiral two-dimensional gauge theory are described by a discrete
random matrix model, the Meixner ensemble, and this has implications for
both systems. (The ferroelectric phase of the six-vertex model with domain wall boundary conditions was already studied using the Meixner ensemble in ~\cite{Bleher}.) In particular, it will allow us to derive a unitary matrix model
description for chiral Yang-Mills theory on $S^{2}$ and also to give a
stochastic interpretation of the partition function, related to the
one given in~\cite{Szabo:2010sd}. Another novel consequence is that its partition function satisfies the Painlev\'{e}~V equation. These
results also apply to the six-vertex model in the ferroelectric phase,
but in that case one must always consider a certain limit of the
model, as we explain later on.

\subsection{Gauge theory characterization of phases in the six-vertex model}

Following~\cite{Bleher,Disor,Anti} we study the
asymptotic large $N$ behaviour of the free energy of the six-vertex model with domain
wall boundary conditions for all three of its phases: the ferroelectric
phase (which is related to chiral Yang-Mills theory on $S^2$), the
disordered phase, and the antiferroelectric phase. Recall that the usual large $N$ expansion of the partition function $Z_{N}$
follows the topological expansion%
\begin{equation}
F_{N}=\log Z_{N}=\sum_{g=0}^{\infty }\, F_{g}(t)\, N^{2-2g} \label{topological}
\end{equation}%
introduced by 't Hooft \cite{'tHooft:1973jz} and studied later on in the
context of matrix models \cite{Brezin:1977sv,Bessis:1980ss}. In particular,
in \cite{Bessis:1980ss} a model with a quartic potential was shown to
follow the asymptotic expansion (\ref{topological}), and more recently the topological expansion has
been rigorously proved for matrix models with polynomial potentials in
\cite{proof}. These matrix models lead to a determined moment
problem~\cite{review}, but matrix models with weaker confining potentials
behave very differently. In particular, Chern-Simons gauge theory
is described by matrix models whose weight function leads to an
undetermined moment problem \cite{Tierz,dht}, and indeed the related large
$N$ expansion of closed topological string theory contains terms that do not appear in (\ref%
{topological})~\cite{Marino(05)}.

As we show in the following, the disordered and
antiferroelectric phases of the six-vertex model with domain wall boundary
conditions exhibit a behaviour that goes beyond (\ref{topological}) and that
is typical of simple gauge theories, such as two-dimensional Yang-Mills theory
or Chern-Simons theory. In the disordered phase, for example, we find a term of the form $%
\kappa \, \log N$, which is absent in (\ref{topological}) but present in Chern-Simons
theory on the three-sphere $S^3$ even in the semiclassical limit. The antiferroelectric phase, on the other hand,
exhibits an oscillatory behaviour due to a term involving a theta-function; we shall see that this
behaviour is characteristic of a matrix model with a multi-cut solution and
appears in the large $N$ expansion of Chern-Simons gauge theory on a lens space.

The $SU(N)$
Chern-Simons gauge theories that we consider in this paper are equivalent~\cite{Beasley:2005vf,Caporaso:2005ta,Blau:2006gh,Griguolo:2006kp} at level $k\in\IZ$ to $q$-deformed Yang-Mills theory
on $S^2$, with the identification of the string coupling
$g_s= \frac{2\pi {\,\mathrm{i}\,}}{k+N}$ and
$$
q:={\,\mathrm{e}}\,^{-g_s} \ .
$$
This equivalence was exploited in the context of certain
one-dimensional integrable models
in~\cite{Szabo:2010qv}.
In particular, the disordered
phase is closely related to the full (non-chiral) \emph{topological}
Yang-Mills theory on $\Sigma= S^{2}$; in this limit the
two-dimensional gauge theory partition function computes
the
symplectic volume of the corresponding moduli space of flat
connections~\cite{Witten:1991we}. This is to be compared with the physically simpler
ferroelectric phase, which is related to the chiral sector of the
non-topological gauge theory that computes the orbifold Euler
characters of Hurwitz spaces.

\begin{figure}[h]
\bigskip
\begin{center}
\epsfxsize=2 in\epsfbox{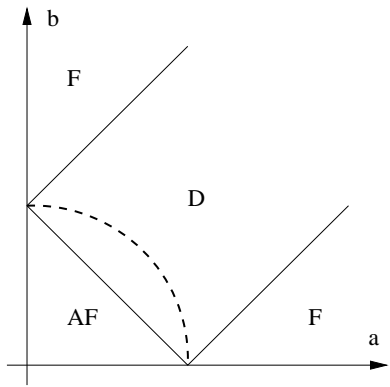}
\end{center}
\caption{Phase diagram for the six-vertex model with domain wall
  boundary conditions, depicting the phase boundaries separating the
  ferroelectric~(F), disordered~(D), and antiferroelectric~(AF)
  regimes as functions of the Boltzmann weights $a$ and $b$ (with
  $c=1$). The dashed circular arc denotes the free fermion line.}
 \label{phase_diagram1}
\end{figure}

Fig.~\ref{phase_diagram1} depicts the standard phase diagram for the six-vertex model with domain wall boundary conditions.
Tab.~\ref{phase_gauge} illustrates the analogous behaviour of gauge theory for each of the three phases. These correspondences with Chern-Simons theory or two-dimensional Yang-Mills theory are not exhibited by other vertex models.
\begin{table}[h]
\bigskip
\begin{tabular}{||c||c||c||c||c||}
\hline\hline
{\footnotesize \bf Phase} & ${\scriptstyle \mbf{Z_{N}}}$ & {\footnotesize \bf Gauge theory} & {\footnotesize \bf String
interpretation} & {\footnotesize \bf Matrix model matches} \\ \hline\hline
{\footnotesize F} & ${\scriptstyle G^{N}\, F^{N^{2}}}$ & {\footnotesize Chiral} & {\footnotesize Branched covers} & {\footnotesize Yes} \\ \hline\hline
{\footnotesize D} & ${\scriptstyle N^{\kappa }\, \e^{N^{2}\,f}}$ & {\footnotesize %
Topological}& {\footnotesize Flat connections} & {\footnotesize Large} ${\scriptstyle N}$\\ \hline\hline
{\footnotesize AF} & ${\scriptstyle \theta_{4}(N\, \omega )\, F^{N^{2}}}$ & ${\scriptstyle q}${\footnotesize-deformed} & {\footnotesize Topological strings} & {\footnotesize Large} ${\scriptstyle N}$ \\ \hline\hline
\end{tabular}%
\bigskip
\caption{Gauge theory characteristics in each phase of the six-vertex model depicted in fig.~\ref{phase_diagram1}.}
\label{phase_gauge}\end{table}
Note that the only  case where the deviation from the usual topological
expansion of matrix models with polynomial potentials does not appear
is in the ferroelectric phase, which is exactly the case where there is a more precise match in terms of the associated matrix models. It would be very appealing to relate the different gauge theory behaviours corresponding to the different phases in terms of the spontaneous symmetry breaking from a disordered state to an ordered state, for example in the breaking of the topological symmetry of the gauge theory of the disordered phase via the appearance of a Yang-Mills coupling in the chiral gauge theory of the ferroelectric phase.

\subsection{Outline}

The remainder of this paper is organized as follows. In \S\ref{YM2} we
describe the partition function of chiral Yang-Mills theory on $S^{2}$ from the perspective of
random matrix theory, identifying the pertinent classical random matrix ensemble as the
Meixner ensemble.
We interpret ${Z}_{\mathrm{YM}}^{+}\left( S^2\,,\,SU(N)\right) $ as a
particular case of the normalization of the $z$-measure, which unravels the integrability
properties of the chiral sector of the gauge theory; for example, it
shows that the partition function is related to a tau-function of the Painlev\'{e}~V
transcendent. Using the expressions for the partition function
involving Toeplitz determinants and the Schur measure,
following~\cite{Szabo:2010sd} we show that there is an alternative
matrix model description to the usual discrete matrix
models~\cite{Gross:1992tu,Gross:1994mr} based on $n\times n$ unitary
matrix models; as in~\cite{Szabo:2010sd} there is also
a natural stochastic
interpretation of the chiral gauge theory. In \S\ref{6V} we explain basic facts about the homogeneous six-vertex model with domain wall boundary conditions for each of its phases in the thermodynamic limit. We obtain the explicit mapping between the partition function of the ferroelectric phase and that of chiral Yang-Mills theory on $S^2$ in the large $N$ limit, and describe various features of this correspondence. We also suggest that the general behaviour of the ferroelectric phase can be captured by generalized two-dimensional Yang-Mills theory, and relate this to the characterization of the partition function as a tau-function of an integrable Toda lattice hierarchy. We further argue that the disordered and antiferroelectric phases are described by Chern-Simons gauge theory on the three-sphere $S^3$ and the lens space $L(2,1)=S^3/\IZ_2$ respectively (equivalently particular $q$-deformations of Yang-Mills theory on $S^2$), and describe various properties of the mappings in this case. In Appendix~A we summarize some technical details of the relationship between the Gross-Witten model and Painlev\'e equations.

\subsection*{Acknowledgments}

We thank R.~Weston for helpful discussions. The work of RJS was supported by grant ST/G000514/1 ``String Theory
Scotland'' from the UK Science and Technology Facilities Council. The work of MT has been supported by a Lady Davis 
fellowship at the Hebrew University of Jerusalem and by the project "Probabilistic approach to finite and infinite 
dimensional dynamical systems" (PTDC/MAT/104173/2008) at the Universidade de Lisboa.

\section{Chiral Yang-Mills theory and the Meixner ensemble\label{YM2}}

\subsection{Meixner matrix model\label{subsec:Meixner}}

We begin by deriving the discrete matrix model representation of the chiral partition
function $Z_{\rm YM}^{+}\left( \mathrm{\Sigma },SU(N)\right) $. The
irreducible representations $\lambda$ of $SU(N)$ correspond to Young
diagrams for which the
number of non-zero rows $n$ satisfies the constraint $n\leq N$. The row lengths are denoted by
$\lambda _{i}$
for all $i=1,\dots,n$ and they define a partition
$\lambda=(\lambda_1,\dots,\lambda_n)$,
i.e. $\lambda_i\geq\lambda_{i+1} \geq 0$. The quadratic Casimir
operator in these
variables is%
\begin{eqnarray*}
C_{2}(\lambda ) = N\, \sum_{i=1}^{n}\, \lambda
_{i}+\sum_{i=1}^{n}\, \lambda
_{i}\, \left( \lambda _{i}+1-2i\right) \ ,
\end{eqnarray*}
while the representation dimensions are given by
\begin{eqnarray*}
\dim\lambda= \frac{|\lambda|\,
  !}{\prod\limits_{i=1}^N\,(\lambda_i-i+N)!} \ \prod_{1\leq i<j\leq
  N}\, (\lambda_i-\lambda_j+j-i)
\end{eqnarray*}%
where $|\lambda|=\sum_i\, \lambda_i$ is the number of boxes of
$\lambda$.

If one considers only small representations, whose row lengths are all $<N,$
then the Casimir eigenvalue is linearized in the large $N$ limit. The
discrete matrix model then follows from
the heat kernel expansion (\ref{HK}) by dropping the constraint on the
number of rows and is given by~\cite{Gross:1992tu}
\begin{equation}
Z_{\rm YM}^{+}\big(\Sigma\,,\,SU(N)\big) =\sum_{n_{i}\in
\mathbb{N}
}~ \prod\limits_{i<j}\, \left( n_{i}-n_{j}\right) ^{2-2h} ~
\prod_{i=1}^{N}\, \e%
^{-g_{s}\, n_{i}} \ ,  \label{chiral}
\end{equation}%
where we have rescaled the coupling constant $g_s\, N\to g_s$, which
is held fixed for $N\to\infty$, i.e. we work in the 't~Hooft limit of
the gauge theory. Throughout we drop irrelevant
overall numerical constants.

The full partition function is given by a discrete Gaussian matrix
model~\cite{Gross:1994mr}. In contrast, the
matrix model (\ref{chiral}) is an orthogonal polynomial ensemble~\cite%
{Konig} which appears naturally in the study of stochastic last passage
models~\cite{Johansson1}. Its associated discrete orthogonal polynomials are the
monic Meixner polynomials which are orthogonal with respect to the family of
weights%
\begin{equation}
\omega_{\rm Meix}\left( n;K\right) =\binom{n+K-1}{n}\, q^{n} \ ,
\qquad n\in\IN  \label{weight}
\end{equation}%
parametrized by $K\in\IN$ and $q\in(0,1)$, and which can be expressed
through hypergeometric functions as ${\sf M}_m(n;K)={}_2F_1\big(\stackrel{\scriptstyle -m,-n}{\scriptstyle K}\,\big|\,1-q\big)$; the orthogonality relation is
$$
\sum_{n=0}^\infty\, \omega_{\rm Meix}(n;K)\ {\sf M}_m(n;K)\, {\sf M}_{m'}(n;K)= \frac{\delta_{m,m'}}{\omega_{\rm Meix}(m;K)} \ .
$$
The associated Meixner
ensemble is a discrete Coulomb gas model on $%
\mathbb{N}
$ with joint eigenvalue probability distribution~\cite{Johansson1,Johansson2}%
\begin{equation}
P_{\rm Meix}(n_1,\dots,n_{N};K) = \frac{1}{Z^{\rm Meix}_{N}(K)} \ \prod\limits_{1\leq i<j\leq N}\left( n_{i}-n_{j}\right)
^{2}~ \prod\limits_{j=1}^{N}\,\omega_{\rm Meix} \left( n_{j};K\right) \ .  \label{Meixner}
\end{equation}%
Hence the partition function of the Meixner ensemble $%
Z_{N}^{\rm Meix}(K)$ with $K=1$ and the identification $q=\e^{-g_s}$ in (\ref{weight}%
) coincides with the genus $h=0$ chiral gauge theory partition
function $Z_{\rm YM}^{+}(S^{2},SU(N))$ in (\ref{chiral}).

One virtue of this matrix model formulation is that the Meixner ensemble is more general and defines natural extensions of
the chiral gauge theory defined by (\ref{chiral}). For $K>1$, it
perturbs the linear potential in (\ref{chiral}) by the logarithmic
potential
$$
V_K(n_i)=\sum_{s=1}^{K-1} \, \log\Big(\,\frac{n_i+s}s \, \Big) \ ,
$$
and in this sense it can be regarded as a discrete Penner
matrix model. We will use some instances of this extension later on in the context of generalized two-dimensional Yang-Mills theory~\cite{Douglas:1994pq,Ganor:1994bq}.

\subsection{$z$-measure}

The $z$-measure is a two-parameter family of probability
distributions on integer partitions $\lambda$ (equivalently Young diagrams) that
originally appeared in the harmonic analysis of the infinite symmetric
group~\cite{z-measure93}. It has been understood in further detail more
recently in~\cite{BO1,BO2}. Consider the quantity %
\begin{eqnarray}
\Qsf_{n}^{z,z^{\prime }, q } = \left( 1- q \right) ^{z\, z^{\prime
  }}\, D_n(\sigma)
\label{QsfToeplitz}\end{eqnarray}
defined for $z,z'\in\IC$, $n\in\IN$ and $ q\in(0,1)$ by an $n\times
n$ Toeplitz
determinant
\beq
D_n(\sigma) =
\det_{1\leq i,j\leq n}\, \big[
\sigma_{i-j}\big] \ ,
\label{T}\eeq
where $\sigma_m=\sigma_m(z,z', q)$, $m\in\IZ$ are coefficients in the Fourier series
expansion of the associated symbol function
\begin{eqnarray*}
\sigma(\zeta):= \big( 1+\sqrt{ q }\, \zeta
\big) ^{z}\, \big( 1+\sqrt{ q }\, \zeta^{-1} \big) ^{z^{\prime }} =
  \sum_{m=-\infty }^{\infty }\, \sigma_{m}\, \zeta ^{m} \ , \qquad \zeta\in
  S^1 \ .
\end{eqnarray*}%
A representation theory definition of $\Qsf_{n}^{z,z^{\prime }, q }$ states
that if $z'=\overline z$ then $%
\Qsf_{n}^{z,\overline z, q }$ is the distribution function of the first row of
the random Young diagram distributed according to a $z$-measure~\cite%
{BO1,BO2}. There are a great number of interesting properties associated to this
measure; in particular, it can be expressed as a Fredholm determinant with a hypergeometric
kernel that allows one to establish a very concrete and detailed connection with
the discrete Painlev\'{e}~V equation. We show that a particular case
of the Toeplitz determinant (\ref{QsfToeplitz}) describes $SU(N)$ chiral Yang-Mills
theory on $S^{2}$.

For this, we use Gessel's formula which expresses a certain series in Schur functions in
terms of a Toeplitz determinant. In terms of the Schur measure
parametrized by sets of variables $x=(x_i)_{i\geq1}$ and
$y=(y_i)_{i\geq1}$, the unnormalized probability distribution for the
number of boxes in the first row of a random Young diagram is given by~\cite%
{Szabo:2010sd}%
\begin{equation}
\mathcal{P}_{N}(x,y):= \sum_{\lambda \,:\,\lambda _{1}\leq N}\,{%
\mathfrak{s}}_{\lambda }\left( x\right) \,{\mathfrak{s}}_{\lambda }\left(
y\right) = D_N(A) \ ,  \label{Toeplitz}
\end{equation}%
where $\mathfrak{s}%
_{\lambda }(x):= \det_{i,j} \big(x_{i}^{\lambda_{j}+n-j}\big)\big/ \det_{i,j}
\big(x_{i}^{n-j}\big)$ are the Schur polynomials and %
\begin{equation}
A_{m}=A_{m}(x,y)=\sum_{l=0}^{\infty }\,\mathfrak{e}_{l+m}(x)\,\mathfrak{e}%
_{l}(y)
\end{equation}%
with $\mathfrak{e}_{l}(x)$ the $l$-th elementary symmetric function. The symbol of the Toeplitz determinant $D_N(A)$ in
(\ref{Toeplitz}) is \cite%
{Szabo:2010sd}
\begin{equation}
A\left(\zeta \right) =\sum_{m=-\infty }^{\infty
}\,A_{m}(x,y)\,\zeta^{m}=\prod\limits_{i\geq 1} \,\left(
  1+x_{i}\,\zeta\right) \,\left( 1+y_{i}\,\zeta^{-1}\right)\ .
\end{equation}%
This symbol coincides with that of (\ref{QsfToeplitz}) for $z=z'=N$
and $x_i=y_i=\sqrt{ q}$ for $i=1,\dots,N$, giving an expansion of
(\ref{QsfToeplitz}) in terms of Schur polynomials as
\begin{eqnarray}
&& D_{n}(\sigma)=\sum_{\lambda \,:\,\lambda _{1}\leq n}\,{\mathfrak{s}}_{\lambda
}\big( \, \underbrace{ \sqrt{ q },\ldots ,\sqrt{ q }}_N\, \big) ^{2}= \sum_{\lambda \,:\,\lambda _{1}\leq n}\, q ^{\left\vert \lambda
\right\vert }\, {\mathfrak{s}}_{\lambda }\left( 1,\ldots,1\right)^{2}
= \sum_{\lambda \,:\,\lambda _{1}\leq n}\, q ^{\left\vert \lambda
\right\vert }\, (\dim\lambda)^{2} \ .
\label{Dk}\end{eqnarray}%

We have to take into account that the Schur polynomial is non-zero as long as its number of variables (N in this case) is at least equal 
to the length of the partition. Hence, making the identification $q=\e^{-g_s}$ again, we find
that the chiral gauge theory partition function ${Z}_{\mathrm{YM}%
}^{+}\left(S^2\,,\,SU(N)\right) $ from (\ref{chiral}) is a particular case of the $z$%
-measure%
\begin{equation}
{Z}_{\mathrm{YM}}^{+}\left(S^2\,,\,SU(N)\right) = 
D_N(\sigma) = (1-q)^{-N^2}\, \Qsf_{N }^{N,N,%
q} \ .
\label{limit}\end{equation}%
This connection with the $z$-measure also follows directly from the
identification of the Meixner ensemble as the discrete matrix
model underlying the chiral sector of Yang-Mills theory on $S^2$,
since choosing $z=N$ and $z^{\prime }=N+K-1$ in the
$z$-measure leads to
the Meixner distribution (\ref{Meixner})~\cite{BO3}. In the ferroelectric
phase of the six-vertex model with domain wall boundary
conditions, this was already pointed out in~\cite{Bleher}.

As an application of this result, let us explicitly evaluate ${Z}_{\mathrm{YM}%
}^{+}\left(S^2\,,\,SU(N)\right) $. For this, we use the Cauchy identity for the Schur measure~\cite{Szabo:2010sd}%
\begin{equation*}
\sum_{\lambda }\,\mathfrak{s}_{\lambda }(x)\,\mathfrak{s}_{\lambda
}(y)=\prod\limits_{i,j= 1}^{N}\,\frac{1}{1-x_{i}\,y_{j}} \ .
\end{equation*}%
Then with the specialization $x_{1}=\dots=x_{N}=y_{1}=\dots=y_{N}=\sqrt{ q }=\e^{-g_s/2}$ leading to (\ref{Dk}), we find%
\begin{equation*}
{Z}_{\mathrm{YM}}^{+}\left(S^2\,,\,SU(N)\right) =\big( 1-%
\e^{-g_s}\big)^{-N^{2}} \ .
\end{equation*}%
This is precisely the formula
which was computed in~\cite{Gross:1992tu} in a much more cumbersome way, by
attempting to derive the system of discrete orthogonal polynomials associated to the matrix model; these polynomials are of course the Meixner polynomials. Below we consider various other applications of the representation (\ref{limit}).

\subsection{Double scaling limit\label{DSlimit}}

The $z$-measure contains the Plancherel measure which is the probability
distribution on Young diagrams that appears in the description of
four-dimensional supersymmetric gauge theory in terms of random partitions~\cite%
{NekOuk,LMN}; in this sense the $z$-measure can be regarded as a deformation of the Plancherel measure. Moreover, a certain double scaling limit of the $z$-measure leads to the
Poissonized version of the Plancherel measure. This measure is in turn directly
related to the Gross-Witten model. This
identifies a double scaling limit of the chiral gauge theory in which
it coincides with the one-plaquette reduction of lattice gauge theory
defined in (\ref{latticeint}).

The scaling can be described as follows~\cite{BOStoch}. Generalizing
(\ref{Dk}) we expand (\ref{QsfToeplitz}) into Young diagrams as
\begin{equation}
\Qsf_{n}^{-z,-z^{\prime }, q }=\sum_{\lambda\,:\,\lambda_1\leq n }\, M_{z,z^{\prime }, q }\left(
\lambda \right) \ ,
\label{Qsfzmeas}\end{equation}%
which defines the mixed $z$-measures
\begin{equation} \label{Q2}
M_{z,z^{\prime }, q }\left(
\lambda \right) = (1-q)^{z\,z'}\,\Big(\, \frac{\dim \lambda }{\left\vert
\lambda \right\vert\, !}\, \Big) ^{2}\, q^{|\lambda|}\, \prod_{(i,j)\in\lambda}\, (z+j-i)\,(z'+j-i)
\end{equation}
on partitions (with the product over the boxes of $\lambda$). The Poissonized version of the
Plancherel measure
$\mathcal{M}_{\mathrm{Planch}}(\lambda) =(\dim\lambda)^2/|\lambda|\, !$ is
defined by
\begin{equation*}
\mathcal{M}_{\mathrm{Planch},\theta }\left( \lambda \right) =\mathcal{M}_{%
\mathrm{Planch}}\left( \lambda \right) \,\e^{-\theta }\, \frac{\theta ^{\left\vert \lambda
\right\vert }}{\left\vert \lambda \right\vert\, !}= \e^{-\theta }\, \theta
^{\left\vert \lambda \right\vert }\, \Big(\, \frac{\dim \lambda }{\left\vert
\lambda \right\vert\, !}\, \Big) ^{2} \ , \qquad \theta>0 \ .
\end{equation*}%
Then one has
\begin{equation}
\lim_{\substack{ z,z^{\prime }\rightarrow \infty ,\text{ } q \rightarrow 0
\\ z\, z^{\prime }\, q = \theta }}\, M_{z,z^{\prime }, q }\left( \lambda
\right) =\mathcal{M}_{\mathrm{Planch},\theta }\left( \lambda \right) \ .
\label{scaling}
\end{equation}%

The Poissonized Plancherel measure is intimately related to the Gross-Witten
model~\cite{GW}, which is defined by the unitary one-matrix model with partition function
\begin{equation}
Z_{\rm GW}(\theta)=\int_{U(N)} \, \dd U~\exp \Big(\, \sqrt{\theta }\, \Tr\big( U+U^{-1}
\big)\, \Big) \ .
\label{GW}\end{equation}
Expanding the integrand in $U(N)$ characters and using standard
orthogonality relations gives an expansion of (\ref{GW}) into Young diagrams as~\cite{Bor}
\begin{equation} \label{ZGWPlanch}
Z_{\rm GW}(\theta) =\e^{\theta }\, \sum_{\lambda \,:\,\lambda _{1}\leq N}\, \mathcal{M}_{\mathrm{Planch}%
,\theta }\left( \lambda \right) \ .
\end{equation}%
We can give a natural physical meaning to this result by interpreting the scaling (\ref{scaling}) at the level of chiral Yang-Mills theory on $S^{2}$. It is a large $N$
limit, but the limits $ q \rightarrow 0$ and $z\, z^{\prime }\, q = \theta $ give $q= \theta /N^{2}$,
and hence%
\begin{eqnarray}
N \rightarrow \infty \ , \qquad g_s= 2\log \big( N\big/\sqrt{\theta }\
\big) \ .
\label{scaling-YM}\end{eqnarray}%
This limit is consistent with the fact that the heat kernel
representation (\ref{HK}) is a strong coupling expansion of the
two-dimensional gauge theory. Hence assuming that the scaling
(\ref{scaling}) can be interchanged with the partition sum
in (\ref{Qsfzmeas}), we find that the chiral partition function
${Z}_{\mathrm{YM}}^{+}\left( S^2\,,\, SU(N)\right) $ from
(\ref{limit}) in the double scaling limit (\ref{scaling-YM}) is equal
to the partition function (\ref{GW}) of the Gross-Witten model in the large $N$ limit
\begin{equation}
\lim_{\substack{ N\rightarrow \infty  \\ g_{s}= 2\log (
      N/\sqrt{\theta }\,) }}\,
{Z}_{\mathrm{YM}}^{+}\left(S^2\,,\, SU(N)\right) = \e^{-\theta}\, Z_{\rm GW}(\theta) \ .
\label{doubleZYMGW}\end{equation}

This result implies that the two gauge theories coincide in the $N\to\infty$ limit. It also shows why we need to consider the $z$-measure, even if only a very particular case, in order to understand the connection between ${Z}_{\mathrm{YM}%
}^{+}\left( S^{2},SU(N)\right) $ and Painlev\'e transcendents for any $N$. The dependence on $N$ in that case comes from a very particular specification of the two additional complex parameters $z$ and $z'$ in the $z$-measure. This specification of $z$ and $z'$ to integer values enjoys additional symmetries described in the work~\cite{Oksl2}, where it is shown to imply a connection with the counting of branched covers of $S^2$. This result fits in nicely with the fact that the chiral sector of two-dimensional Yang-Mills theory is related to Hurwitz theory~\cite{Gross:1993hu,Gross:1993yt}; generally the large $N$ expansion of the chiral partition function ${Z}_{\mathrm{YM}%
}^{+}\left( \Sigma,SU(N)\right) $ can be expressed as a Gross-Taylor string series in branched covering maps of the Riemann surface $\Sigma$. Moreover, consideration of the general $z$-measures (\ref{Q2}) can be interpreted as generalized two-dimensional Yang-Mills theories~\cite{Douglas:1994pq,Ganor:1994bq} which are more general than those given by the Meixner ensembles with $K>1$.

\subsection{Unitary matrix models and tau-functions of Painlev\'{e} equations\label{subsec:unitary}}

The expression (\ref{limit}) for the partition function
${Z}_{\mathrm{YM}}^{+}(S^2,SU(N)) $ in terms of a Toeplitz determinant
immediately leads to a unitary one-matrix model of chiral Yang-Mills
theory on $S^2$. This follows from the
Heine-Szeg\H{o} identity~\cite{Szabo:2010sd}
$$
\int_{U(n)}\, \dd U \ \det\sigma(U)=D_{n}\left(\sigma%
\right)
$$
which relates a generic $n\times n$ Toeplitz determinant to the integral of
its symbol over the unitary
group $U(n)$. In the case at hand this leads to
\begin{equation}
{Z}_{\mathrm{YM}}^{+}\left( S^2\,,\,SU(N)\right) =\int_{U(N)}\, \dd U~ \det\big( 1+\sqrt{ q } \ U^{-1}\big)^N
\, \det\big( 1+\sqrt{ q } \ U \big)^N \ .
\label{YMunitary}\end{equation}%

Notice that the rank $N$ of the gauge group also appears as a parameter that modifies the weight function. The use of the more generic parameters $z$ and $z'$ instead of $N$ as powers in (\ref{YMunitary}) should describe a generalized two-dimensional Yang-Mills theory. Another implication of that matrix model description of the partition function is an equivalence between such a matrix integral and the one given by the Meixner ensemble. In the double scaling limit considered in \S\ref{DSlimit}, the matrix model (\ref{YMunitary}) converges to the large $N$ limit of the Gross-Witten model (\ref{GW}).

An immediate consequence of the expression for the chiral partition function in terms of the random matrix average (\ref{YMunitary}) is that, by the work~\cite{ForrWitte}, the partition function ${Z}_{\mathrm{YM}}^{+}(S^2,SU(N))$ is a tau-function of the Painlev\'{e} V equation. A systematic comparison with the Gross-Witten model is also possible. Forrester and Witte found that the two associated matrix integrals are special instances of tau-functions
\begin{equation}
\tau ^{{\rm III}}\left[ N\right] \left( t;\mu \right) =
\int_{U(N)}\, \dd U \ \det\big(U^\mu\big) \ \exp\Big(\mbox{$\frac12$}\, \sqrt t\, \Tr
\big(U+U^{-1}\big) \Big),
\label{GWt}\end{equation}%
and
\begin{equation*}
\tau ^{\rm V}\left[ N\right] \left( t;\mu ,\nu \right) =
\int_{U(N)}\, \dd U \ \det(1+U)^\mu\, \det\big(1+U^{-1}\big)^\nu \ \exp\big(t\, \Tr(U) \big) \ ,
\end{equation*}
for the Painlev\'{e} III and Painlev\'{e} V equations respectively \cite%
{ForrWitte}, in the sense of the Hamiltonian formulation of the
Painlev\'{e} equations (see Appendix~A for details). In both cases one
can identify a B\"acklund transformation that can be used to establish
a Toda chain equation for the corresponding tau-function sequence.

We then have
\begin{equation*}
Z_{\mathrm{GW}}(\theta )=\tau ^{\mathrm{III}}\left[ N\right] \left( 4\theta
;0\right)
\end{equation*}%
and
\begin{equation*}
\lim_{g_{s}\to 0}\, {Z}_{\mathrm{YM}}^{+}\big(S^{2}\,,\,SU(N)\big)=\tau ^{\mathrm{V}%
}\left[ N \right] \left( 0;N,N\right) \ .
\end{equation*}%

Not only can the partition functions of the two Yang-Mills theories be identified with tau-function of Painlev\'{e} equations, but certain averages in the corresponding matrix models also have this property. Averages of determinants in random matrix ensembles often appear in the random matrix theory approach to quantum chromodynamics (see~\cite{Verbaarschot:2005rj} for a review) and, in particular, the matrix integral (\ref{GWt}) appeared in \cite{Leutwyler:1992yt} describing a QCD partition function in the sector of topological charge $\mu$. Hence the partition function $Z_{\mu}$ given in \cite[eq.~(9.25)]{Leutwyler:1992yt} is a tau-function of the Painlev\'{e} III$^\prime$ equation, a property already exploited in \cite{Verbaarschot:2005rj}.

Notice that we have mostly emphasized the connection with Painlev\'{e} equations, whereas it is 
well-known that there exists other relationship with integrable hierarchies, mostly of the 
KP type \cite{Foda}. Tau functions in the Hamiltonian formulation of Painlev\'{e} equations are 
related to tau functions of the Toda lattice \cite{ForrWitte}, so both results are related and we 
expect to explore such a relationship elsewhere.

\subsection{Corner growth model}

The
models considered in this
paper are naturally related to the corner growth model with geometric
weights~\cite{Johansson1}; this model is described in terms of the Meixner ensemble,
and hence is intimately related to the chiral sector of
Yang-Mills theory on $S^{2}$ studied in this section and to the six-vertex model with domain wall
boundary conditions studied in \S\ref{6V}.
Let $\omega \left( i,j\right)$ for $\left( i,j\right) \in
\IN^2$ be independent geometric random variables and define%
\begin{equation*}
G(M,N)=\max_{\pi:(1,1)\mapsto (M,N) }\ \sum_{(i,j)\in \pi}\, \omega \left( i,j\right) \
,
\end{equation*}%
where the maximum is taken over all up/right paths $\pi$ in $\IN^2$ from $(1,1)$
to $(M,N)$. This model can be given several
probabilistic interpretations -- as a randomly growing Young diagram,
as a
totally asymmetric one-dimensional exclusion process, as a certain
directed polymer in a random environment at zero temperature, or as a kind of
first-passage site percolation model.

We
suppose that $%
\omega \left( i,j\right) $ are distributed according to the
probability measure
\begin{equation*}
\mathcal{P}\big[ \omega \left( j,k\right) =m\big] =(1-q)\, q^{m} \ , \qquad m\in\IN \ .
\end{equation*}%
In this model, Johansson proved that the probability for $G(M,N)$ being
smaller than a certain value can be expressed as~\cite%
{Johansson1,Johansson2}%
\begin{equation*}
\mathcal{P}\big[ G(M,N)\leq t\big] =\frac{1}{\mathcal{Z}_{M,N}}\, \prod\limits
_{\substack{ n_i\in
\mathbb{N}
 \\ \max_i n_{i} \leq t+N-1}}\, \left( n_{i}-n_{j}\right)
^{2}~ \prod\limits_{i=1}^{N}\, \binom{n_{i}+M-N}{n_{i}}\, q^{n_{i}} \ ,
\end{equation*}%
and hence the normalization constant of the process
$\mathcal{Z}_{M,N}$ coincides with the
partition function of the Meixner ensemble $Z_N^{\rm Meix}(K)$ for
$K=M-N+1$. In the symmetric case $%
M=N$, this is just the chiral partition function of Yang-Mills theory
on $S^{2},$ i.e. $\mathcal{Z}%
_{N,N}=Z_{\rm YM}^{+}(S^{2},SU(N)).$ This result is consistent with
the double scaling limit (\ref{doubleZYMGW}); if one takes $q=\alpha
/N^{2},$ then $G(N,N)$ converges in distribution to $L(\alpha )$ as
$N\rightarrow \infty $, where $L\left( \alpha \right) $ denotes the
Poissonized version of the
random variable describing the longest increasing subsequence in a
random permutation, whose distribution is
given by the Gross-Witten model.

\section{Gauge theory descriptions of the six-vertex model\label{6V}}

\subsection{Six-vertex model with domain wall boundary conditions}

Let us consider now the structure of the partition function (\ref{ZN6V}) for the six-vertex model.
The domain wall boundary conditions are only defined for square lattices, and demand
that the external horizontal arrows are outgoing while the external
vertical arrows are incoming. In this model one has six types of vertices $%
\left( a_{1},a_{2},b_{1},b_{2},c_{1},c_{2}\right)$, where the subscript 2
refers to an opposite configuration to that of subscript 1. Conservation laws reduce the weights to the homogeneous case $a_i=a$, $b_i=b$ and $c_i=c$, and the partition function depends only on the two parameters $\frac ac $ and $\frac bc$ due to the identity~\cite{Bleher}
$$
Z_N(a_1,a_2,b_1,b_2,c_1,c_2)=c^{N^2}\, Z_N\big(\mbox{$\frac ac,\frac ac,\frac bc,\frac bc,1,1$}\big) \ .
$$
The
parametrization of the Boltzmann weights associated to the vertices is given by%
\begin{equation}
a=\sinh (t-\gamma ) \ , \qquad b=\sinh (t+\gamma ) \qquad \text{and} \qquad c=\sinh (2\gamma
) \ .  \label{BW}
\end{equation}%

The partition function of this model was expressed, using the earlier
work \cite{Kor1}, as a determinant in \cite{IzKor,IzKor2}, and this was used
in \cite{Pz} to find a matrix model expression. The determinant is a Hankel
determinant which has a Hermitian matrix model representation
(in the same way that a Toeplitz determinant has a unitary matrix model
representation \cite{Szabo:2010sd}). The Izergin-Korepin determinant formula is%
\begin{equation}
Z_{N}=\frac{\big(\sinh \left( \gamma -t\right) \, \sinh \left( \gamma +t\right)
\big)^{N^{2}}}{\Big(\, \prod\limits_{n=0}^{N-1}\, n!\, \Big) ^{2}} \ \tau _{N} \ ,
\label{Z}\end{equation}%
where $\tau _{N}$ is the Hankel determinant \cite{Pz}
\begin{equation*}
\tau _{N}(t)=\det_{1\leq i,j\leq N}\,\left[ \frac{\dd^{i+j-2}}{\dd t^{i+j-2}}\phi (t)%
\right]
\end{equation*}%
with
$$
\phi (t)=\frac{\sinh (2\gamma )}{\sinh (t+\gamma )\, \sinh (t-\gamma )} \ .
$$
The
partition function $\tau _{N}$ is a tau-function of the Toda chain hierarchy; in particular, it satisfies the Toda equation
$$
\tau _{N}\, \tau _{N}^{\prime \prime }-(\tau _{N}^{\prime })^2=\tau _{N+1}\, \tau
_{N-1} \ .
$$

To characterize the three phases of the model in the thermodynamic limit $N\to\infty$ one introduces the parameter%
\begin{equation}
\Delta =\frac{a^{2}+b^{2}-c^{2}}{2\,a\, b} \ . \label{phase}
\end{equation}%
The ferroelectric phase is the region $\Delta >1$, the antiferroelectric phase is $\Delta
<-1$, and the disordered phase corresponds to $-1<\Delta <1$. The free fermion curve is given by $%
\Delta =0$ in the space of parameters. The large $N$ asymptotics of the partition function in each phase can be computed by means of the Riemann-Hilbert method~\cite{Bleher,Disor,Anti} and is summarized as follows:
\begin{itemize}
\item[{\bf (F)}] The ferroelectric phase is the region where the two parameters satisfy $\left\vert \gamma \right\vert <t$, and with $\gamma >0$ for any $\varepsilon >0$ as $%
N\rightarrow \infty $ one has~\cite{Bleher}
\begin{equation*}
Z_{N}=C\, G^{N}\, F^{N^{2}}\, \left( 1+\mathcal{O}\big( \e^{-N^{1-\varepsilon
}}\big) \right) \ ,
\end{equation*}%
with $C=1-\e^{-4\gamma},$ $G=\e^{\gamma -t}$ and $F=\sinh \left( t+\gamma
\right)$.
\item[{\bf (D)}] For the disordered phase one has $|t|<\gamma$ and~\cite{Pz,KorPz}
\begin{equation}
Z_{N}=c\, N^{\kappa }\, \e^{ N^{2}\, f}\, \big(1+\mathcal{O}(
N^{-1-\varepsilon }) \big ) \ , \qquad \varepsilon >0 \ ,  \label{Z6V}
\end{equation}%
where $c>0$ is a constant, while %
\begin{equation*}
\kappa =\frac{1}{12}-\frac{2\gamma ^{2}}{3\pi \left( \pi -2\gamma \right) } \qquad \mbox{and} \qquad
f=\log \Big( \, \frac{\pi \, \big( \cos \left( 2t\right) -\cos \left( 2\gamma
\right) \big) }{4\gamma\, \cos \big( \frac{\pi \,t}{2\gamma }\big) }%
\, \Big) \ .
\end{equation*}%
\item[{\bf (AF)}] The remaining antiferroelectric phase is the region $|t|<\gamma$ of parameter space and the $N\rightarrow \infty $ limit of the partition function is given by~\cite{Anti}
\begin{equation}
Z_{N}= c\,\theta _{4}\big(N\,(1+\mbox{$\frac t\gamma$}),q\big) \, F^{N^{2}}\, \big(1+\mathcal{%
O}(N^{-1}) \big) \ , \label{Anti-Z}
\end{equation}%
where $c>0$ is a constant and $F$ is a function on the two-dimensional parameter space given by
\begin{equation*}
F=\frac{\pi\, \sinh (\gamma -t)\, \sinh (\gamma +t)\, \theta _{1}^{\prime }(0,q)}{%
2\gamma \, \theta _{1}\big(1+\mbox{$\frac t\gamma$},q\big) } \ ,
\end{equation*}%
while the elliptic nome $q$ of the theta-function%
\begin{equation*}
\theta _{4}(z,q)=1+2\,\sum_{n=1}^{\infty }\, (-1)^{n}\, q^{n^{2}/2}\, \cos \left(n\, z\right)
\end{equation*}%
is given by $q=\e^{-\pi ^{2}/\gamma }$.
\end{itemize}

\subsection{Ferroelectric phase and chiral Yang-Mills theory}

The match between the six-vertex model and chiral Yang-Mills theory on $S^2$ occurs in the ferroelectric
phase with $\Delta>1$ (and $0<|\gamma|<t$). In this regime the partition function $\tau_N$ can be written as a discrete matrix
model~\cite{Pz}%
\begin{equation}
\tau _{N}=2^{N^{2}}\, \sum_{l_{i}\in\IN}~ \prod\limits_{1\leq i<j\leq N}\, \left( l_{i}-l_{j}\right) ^{2}\,\e^{-2t\,\sum_{i}\, l_{i}} \ \prod\limits_{i=1}^{N}\, \sinh \left( 2\gamma \,
l_{i}\right) \ . \label{fullmodel}
\end{equation}%
Instead of studying this model, in \cite{Pz} the eigenvalues are rescaled as $%
n_{i}=l_{i}/N$ and then exponentially small contributions for large $N$
are neglected by using $\sinh \left( 2\gamma \, n_{i}\, N\right) = \e^{2\left\vert \gamma \right\vert\, n_{i}\, N}/2+\mathcal{O}(\e^{-N})$, leading to the
matrix model%
\begin{equation}
\tau _{N}= (2N)^{N^{2}-N}\, \sum_{n_{i}\in \mathbb{N}
/{N}} \ \prod\limits_{i<j}\, \left( n_{i}-n_{j}\right) ^{2} \, \e^{-2N\, (t-\left\vert \gamma \right\vert )\, \sum_{i}\, n_{i}} +\mathcal{O}(\e^{-N}) \ ,
\label{gammalarge}
\end{equation}%
which is a Hermitian Meixner ensemble.

To compare this matrix model directly to (\ref%
{Meixner}), we want to have a normal scaling of the eigenvalues, i.e. $%
n_{i}\in\mathbb{N}$. This implies that the approximation in \cite{Pz} then requires $\gamma $ to be very large, if we do not
consider the large $N$ limit. Since $|\gamma|<t$ in the ferroelectric phase, we must also have $t$ large such that $t-|\gamma|>0$. Hence for $\gamma\to\infty$ the partition function of the six-vertex model can be mapped to the partition function of chiral $SU(N)$ Yang-Mills theory on $S^2$ via the identification
\begin{equation}
g_s=2\big(t-|\gamma| \, \big) \ .
\label{gs2tgamma}\end{equation}
To write this partition function as a tau-function of the Painlev\'e~V equation, we must further work in the weak coupling regime; this is the requirement that $t-|\gamma|$ must be very small. In this limit, the Boltzmann weights $b$ and $c$ in (\ref{BW}) are very large while $a=t-|\gamma|\to0$; this appears to lead to a reduction to a four-vertex model.

These results also suggest a connection between the limiting expression (\ref{gammalarge}) for the
partition function in the ferroelectric regime and free fermion systems.
Yang-Mills theory on $S^{2}$ can be regarded as a theory of free fermions on a
circle~\cite{cordesmoore}. In~\cite{Minahan:1993tp} the full (non-chiral) partition function on the sphere
is found by computing
the scalar product of the wavefunction for $N$ fermions at position $x=0$ on
the circle at time $t=0$ with the wavefunction of $N$ fermions at $x=0$ at time $t=T$. The chiral part of the Hilbert space $\mathcal{H}_+$ in (\ref{fact}) is the left-moving sector of a $c=1$ conformal field theory~\cite%
{cordesmoore}. This implies that there is a free fermion Fock space $%
\mathcal{H}_+$ whose partition function coincides with the partition function
of the homogeneous six-vertex model with domain wall boundary conditions in
the ferroelectric phase $\Delta >1$ in the limit of~\cite{Pz}. Other relationships between free
fermion models and the six-vertex model with domain wall boundary conditions are known to hold in the
disordered phase~\cite{CP}.

\subsection{Ferroelectric phase and generalized Yang-Mills theory\label{subsec:genYM}}

It is natural interpret the exact result (\ref{fullmodel}) in the context of generalized two-dimensional Yang-Mills theory~\cite%
{Douglas:1994pq,Ganor:1994bq}, whose heat kernel expansion on a Riemann surface $\Sigma$ is given generically by
\begin{equation*}
\mathcal{Z}_{\rm M}^{\rm gen}(g_s,t)=\sum_{\lambda }\, \left( \dim \lambda \right)
^{2-2h} \, \exp
\Big( -g_{s}\,\sum_{p>0}\, t_p\, C_{p}(\lambda ) \Big) \ ,
\end{equation*}
where
$$
C_p(\lambda)=\sum_{i=1}^n\, (\lambda_i-i+1)^p \ \prod_{j\neq i}\, \Big(1-\frac1{\lambda_i-\lambda_j+j-i}\, \Big)
$$
is the $p$-th Casimir operator eigenvalue in the representation $\lambda$. Higher Casimir operators in the heat kernel expansion correspond to higher powers of the field strength $F$ in the gauge theory action (\ref{continuum})~\cite{Ganor:1994bq}. Indeed, as pointed in~\cite{Witten:1991we}, the distinctive properties of invariance under area-preserving diffeomorphisms, the absence of propagating degrees of freedom, and exact self-similarity~\cite%
{Migdal75} are not unique to the two-dimensional gauge theory based on
the Yang-Mills action~(\ref{continuum}).
In the correspondence between Yang-Mills theory on $S^2$ and the six-vertex model, we are led to consider an additional potential $V(l):=\log(2\sinh 2\gamma\, l)$. At strong coupling $t>\gamma\gg1$, this potential has an expansion
$$
V(l)=2\gamma\,l - \sum_{n\geq1}\, \frac1n\, \e^{-4\, n\,\gamma\,l} \ .
$$
Since any polynomial in the Young tableaux weights $\lambda_i$ can be written as a linear combination of Casimir invariants $C_p(\lambda)$, in this regime the partition function of the six-vertex model can be mapped to a modification of chiral Yang-Mills theory by infinitely many higher Casimir operators.

Different truncations of this power series expansion could also be studied approximately using the Meixner ensemble with more general (not necessarily integer-valued) parameter $K$. This is particularly useful if we recall the integrability structure of the model, i.e. that (\ref{fullmodel}) is the tau-function of the Toda chain hierarchy. Tau-functions of the 1-Toda and 2-Toda lattice hierarchies have expansions in terms of
Schur polynomials that are useful in illustrating how the heat kernel expansion of
this generalized two-dimensional Yang-Mills theory compares with the ordinary one in (\ref{HK}) (or (\ref%
{chiral}) for the chiral case). In our particular case, the squared Vandermonde determinant structure implies that the
expansion in terms of products of pairs of Schur polynomials in its diagonal form is the one
required (see \cite{Szabo:2010sd} and references therein). More precisely one has
\begin{equation*}
\tau _{N}\left( t\,,\,\overline{t}\,\right) =\sum_{\lambda }\,c_{\lambda,N}\,%
\mathfrak{s}_{\lambda }(t)\,\mathfrak{s}_{\lambda}(-\overline{t}\,)\ ,
\end{equation*}%
where the coefficients $c_{\lambda,N}$ are Pl\"{u}cker coordinates of an
infinite-dimensional flag manifold and are given as determinants. Note that $c_{\lambda,N}=1$ in the case of the usual chiral two-dimensional Yang-Mills theory; in this way one can interpret the term $\prod_i\, \sinh \left(
2\gamma\, l_{i}\right) $ in (\ref{fullmodel}) geometrically in terms of Pl\"{u}cker coordinates.

The expression (\ref{fullmodel}) exactly describes the
six-vertex model away from the free fermion curve $\Delta =0$. Using Meixner polynomials, a comparative asymptotic study of the relationship between
the exact partition function (\ref{fullmodel}) and the limiting partition function (\ref{gammalarge}) is carried out in \cite%
{Bleher}. This raises the question as to whether or not there is any special connection between the general theory and a free fermion system; the same question
can be applied to the general Meixner ensemble (\ref{Meixner}). In the context of generalized two-dimensional Yang-Mills theory, this question has been answered affirmatively in~\cite{Dubath:2002qv}. In principle one obtains in this way a description of the six-vertex model in the thermodynamic limit from the point of view of Hurwitz theory~\cite{Ganor:1994bq,Dubath:2002qv}.

\subsection{Ferroelectric phase and BF-theory}

As a warm-up to the gauge theory descriptions of the other phases of the six-vertex model with these boundary conditions, we can alternatively
characterize the gauge theory of the ferroelectric phase as a BF-theory on $S^{2}$ with a non-zero theta-angle. This is reminiscent of recent studies of
topological order in condensed matter systems and statistical mechanics~\cite%
{Ardonne:2003wa}, although the connection between a topological gauge theory
and a statistical mechanics model (or its one-dimensional quantum mechanics
counterpart: the XXZ model) is of a different nature.
As discussed in \S\ref{subsec:genYM}, the quadratic Casimir invariant $C_{2}\left(
\lambda \right) $ in (\ref{HK}) can be replaced by any other function of partitions $%
\lambda=(\lambda_1,\dots,\lambda_n) $ since its appearance is due to the fact that one is trying to
reproduce a continuum gauge theory whose action (\ref{continuum}) is quadratic in
the field-strength. For example, instead of the Yang-Mills term $\Tr F^{2}$ in
the chiral sector we can consider a flux term $\Tr F $ in the non-chiral theory that describes $U(N)$ topological Yang-Mills theory in two-dimensions, and is equivalent to a
theta-angle. If we consider the limit $g_s=0$ and a non-zero theta-angle $%
\theta \neq 0$, then the heat kernel expansion (\ref{HK}) for $\Sigma =S^{2}$ and $U(N)$ gauge group reads
\begin{equation*}
\mathcal{Z}_{\rm M}^{\rm top}=\sum_{\lambda }\, \left( \dim \lambda \right) ^{2}\, \exp \big(
-\theta\, C_{1}(\lambda) \big) \ ,
\end{equation*}%
with $C_{1}(\lambda )=\sum_{i}\, (\lambda_{i}-i+1)$ the linear Casimir invariant of the $U(N)$ representation $\lambda$. This leads to the Meixner matrix model
(\ref{chiral}) with the theta-angle as the coupling constant.

Two-dimensional Yang-Mills theory has an alternative description in terms of a BF-theory~\cite{Witten:1991we,cordesmoore}. For $g_s=0$ and $\theta\neq0$, this is ordinary BF-theory which is a pure topological gauge theory with action
\begin{equation*}
S_{\rm BF}=-\ii\int_{\Sigma } \, \Tr\left( B \, F\right) +\theta\,
\int_{\Sigma } \, \Tr\left( B \, K\right) \ ,
\end{equation*}%
where $B$ is a $\mathfrak{u}(N)$-valued zero-form and $K$ is the unit area form on $\Sigma =S^{2}$. This theory is equivalent to the semiclassical limit $\left( k\rightarrow \infty
\right) $ of Chern-Simons gauge theory on $S^{1}\times \Sigma$ with a theta-angle. A quick way of
seeing this is to notice that in this limit the heat kernel expansion (\ref%
{HK}) is just the sum over all dimensions of $U(N)$ representations and, by Verlinde's formula, this
is the $k\rightarrow \infty $ limit of the Chern-Simons partition
function~\cite{Blau:1993tv}.

\subsection{Disordered phase and Chern-Simons theory on $S^3$\label{Disordered}}

The path integral for $SU(N)$ Chern-Simons gauge theory on an oriented compact three-manifold $M$ localizes onto a sum over contributions from flat connections. When $M\to\Sigma$ is a Seifert fibration, the localized gauge theory is equivalent to $q$-deformed Yang-Mills theory on the base Riemann surface $\Sigma$ after analytic continuation of the respective coupling constants. On the three-sphere $M=S^{3}$, there is only the trivial flat gauge connection up to isomorphism, and the partition function reads
\begin{equation*}
Z_{\rm CS}\big(S^3\,,\,SU(N) \big)=\frac{1}{\left( k+N\right) ^{N/2}}\ \prod\limits_{1\leq i<j\leq N}\, 2\sin \Big(\,
\frac{\pi \left(j-i\right) }{k+N} \, \Big) \ .
\end{equation*}%

This partition function can be written as a matrix integral~\cite{Marino(05)}
\beq
&& Z_{\rm CS}\big(S^3\,,\,SU(N) \big)= \frac1{N!\, (k+N)^{N/2}}\, \int_{\IR}\, \prod_{i=1}^N\, \frac{\dd x_i}{2\pi}\ \e^{-x_i^2/2g_s}\ \prod_{i<j}\, \Big(\, 2\sinh\frac{x_i-x_j}2\,\Big)^2
\label{ZCSS3MM}\eeq
with $g_s=\frac{2\pi\ii}{k+n}$. In this description of Chern-Simons theory in terms of matrix models, the computations are carried out 
with q real and it is possible to make contact with Chern-Simons theory by simply identifying 
$g_s=\frac{2\pi\ii}{k+n}$ ~\cite{Tierz}. The orthogonal polynomials of this matrix ensemble are the Stieltjes-Wigert polynomials~\cite{Tierz,DTierz}. This Hermitian matrix model is equivalent to the discrete matrix model~\cite{dht,Szabo:2010qv}
$$
Z_{\rm CS}\big(S^3\,,\,SU(N) \big)=\sum_{n_i\in\IZ}\, \e^{-\frac{g_s}2\, \sum_i\, n_i^2}\ \prod_{i<j}\, \Big(\, 2\sinh\frac{g_s}2\, (n_i-n_j)\, \Big)^2
$$
which defines the partition function of the corresponding $q$-deformed Yang-Mills theory on $S^2$, with the three-sphere regarded as the Hopf fibration $S^3\to S^2$. There is also an $N\times N$ unitary matrix model equivalent whose orthogonal polynomials are the Rogers-Szeg\H{o} polynomials~\cite{DTierz}.

The corresponding free energy can be suitably expressed in terms of
non-perturbative and perturbative contributions
\begin{equation*}
F_{\rm CS}=\log Z_{\rm CS}=F_{\rm np}+F_{\rm p} \ .
\end{equation*}%
The splitting of the exact Chern-Simons free energy into perturbative and
non-perturbative pieces has a physical interpretation in Type~IIA
superstring theory.
The non-perturbative contribution $F_{\rm np}$ is the logarithm of the measure factor
in the path integral, which is not captured by Feynman diagrams, and it gives the exact Chern-Simons partition function in the semiclassical limit $k\rightarrow \infty$~\cite%
[eq.~(2.8)]{Ooguri:2002gx}. It has the
explicit expression%
\begin{equation*}
F_{\rm np}=\log \Big(\, \frac{\left( 2\pi\, g_{s}\right) ^{N^{2}/2}}{%
\mathrm{vol}\big(SU(N)\big)}\, \Big) \ .
\end{equation*}%
The volume of the gauge group is inversely proportional to the Barnes double gamma-function
\beq
G_{2}(N+1)=\prod\limits_{n=0}^{N-1} \, n!
\label{Barnes}\eeq
which has the asymptotic large $N$ expansion~\cite%
{Ooguri:2002gx}
\beq
\log
G_{2}(N+1)&=&\frac12\, N^{2}\, \log N-\frac{1}{12}\, \log N-\frac{3}{4}\, N^{2}
+\frac N2\, \log 2\pi +\zeta ^{\prime }(-1) \nonumber \\ && +\, \sum_{g=2}^{\infty }\, \frac{B_{2g}}{%
2g\, (2g-2)}\, N^{2-2g} \ ,
\label{n!expansion}\eeq
where $\zeta(z)$ is the Riemann zeta-function and $B_{2g}$ are Bernoulli numbers. The full partition function, including the perturbative contribution, is a $q$-deformation of the Barnes $G$-function (\ref{Barnes}) (and hence of ${\rm vol}(SU(N))$).

This leads to the explicit expansion \cite%
{Periwal:1993yu,Ooguri:2002gx}%
\begin{eqnarray}\label{FNP}
F_{\rm np}=\frac{N^{2}}{2}\, \Big( \log g_{s}-\frac{3}{2}%
\, \Big) -\frac{1}{12}\, \log N+\zeta ^{\prime }\left( -1\right) +\sum_{g=2}^{\infty }\, \frac{B_{2g}}{2g\, (2g-2)}\, N^{2-2g} \ ,
\end{eqnarray}%
where again here $g_{s}\, N\to g_s$ is the 't~Hooft coupling which is kept fixed
for $N\rightarrow \infty$. The free energy (\ref{FNP}) has a very precise meaning in terms of closed topological string theory on
the resolved conifold geometry~\cite{Ooguri:2002gx}. It is given by a sum over Riemann surfaces in the pure Coulomb phase, i.e. with a single hole covering the whole worldsheet, and it computes the Euler characters of the
moduli spaces of genus $g$ Riemann surfaces. See~\cite{Ooguri:2002gx} for equivalent string theory and gauge theory interpretations.

The free energy is thus of the type (\ref{Z6V}) which describes the thermodynamics of the six-vertex model with domain wall boundary conditions in the disordered phase. Comparing (\ref{FNP}) and (\ref{Z6V}), we see that both free energies coincide
and the ``unconventional'' factor $N^{\kappa }$, absent in the ferroelectric phase, is given by $\kappa =-\frac1{12}$,
which yields $$\gamma =\mbox{$\frac\pi4$}\, \big(\sqrt{5} -1\big) \ . $$
Note
that $0<\gamma<\frac\pi2$, as required in the disordered phase. The
remaining parameter $t$ is then fixed by the value of the 't Hooft coupling constant through $\frac{1}{2}\,\big( \log g_s-\frac{3}{2}\big)=f$, giving
$$
g_{s}=\e^{3/2}\,\Big(\,\frac{\pi \,\big(\cos \left( 2t\right)
-\cos \left( 2\gamma \right) \big)}{4\,\gamma \cos \big(\frac{\pi \,t}{%
2\gamma }\big)}\,\Big)^{2}\ .
$$

Thus depending on $|t|<\gamma$ and hence on the value of 't Hooft coupling $g_s$, the
vertex model lives at different points on the phase diagram within the
disordered phase. This is an especially interesting phase, because together with the antiferroelectric phase the
important ``artic circle'' phenomena occurs in this
phase -- there are macroscopically big frozen and random domains in typical
configurations, separated in the limit $N\rightarrow \infty $ by an
\textquotedblleft arctic curve\textquotedblright. In the ferroelectric phase, we already know that the term $N^{\kappa }$ is
absent and that there is a correspondence with two-dimensional Yang-Mills theory.
The critical line in the phase diagram of fig.~\ref{phase_diagram1} between the
ferroelectric and disordered phases has $\kappa =\frac14$, while the critical line
between the disordered and antiferroelectric phases has $\kappa =\frac1{12}$ which is the
right factor but with the wrong sign. Note that the error term in (\ref{Z6V}) implies that there is no $\frac1N$ contribution to the free energy, which is consistent with the large $N$ expansion (\ref{FNP}).

The Barnes $G$-function (\ref{Barnes}) also appears in the denominator of the Izergin-Korepin formula (\ref{Z}). It further arises as the partition function of the Gaussian matrix model
\begin{equation*}
\frac1{N!}\, \int_{\IR }\ \prod\limits_{i=1}^{N} \, \dd x_i \ \e^{-a\,x_i^2}\ \prod_{i<j}\, (x_i-x_j)^2 =\left( 2\pi \right)
^{N/2}\, (2a)^{-N^{2}/2}\, G_2(N+1) \ .
\end{equation*}%
Thus the semi-classical Chern-Simons gauge theory on $S^3$ is essentially a Gaussian matrix model, which matches the leading behaviour of the six-vertex model.

On the other hand, the partition function of the disordered phase is not quite a Gaussian matrix
model, which would provide the exact match with the
non-perturbative Chern-Simons partition function, but rather
it comes from a fluctuating model.
The partition function in the disordered
phase is described by a continuous matrix model~\cite{Pz}
\begin{equation}
\tau _{N}=\frac{1}{N!}\, \int_{\IR}\ \prod\limits_{i=1}^{N} \, \dd\lambda_i \ \e^{t\, \lambda _{i}}\, \frac{%
\sinh \frac{\lambda _{i}}{2}\left( \pi -2\gamma \right) }{\sinh \frac{%
\lambda _{i}}{2}\, \pi } \ \prod_{i<j}\, (\lambda_i-\lambda_j)^2 \ . \label{MP}
\end{equation}%
This ensemble can be studied with the monic Meixner-Pollaczek orthogonal polynomials~\cite{linear,freeline}; it should be compared with the matrix model (\ref{ZCSS3MM}). The full
partition function is given by~(\ref{Z}).

On the free fermion line that crosses
both the antiferroelectric and disordered regions in fig.~\ref{phase_diagram1}, characterized by the parameters $\gamma =\frac{%
\pi }{4}$ and $\left\vert t\right\vert <\frac{\pi }{4}$, the full partition
function is just~\cite{freeline}%
\begin{equation*}
Z_{N}\big|_{\gamma=\pi/4}=1 \ .
\end{equation*}%
This means that the matrix model%
\begin{equation*}
\tau_N\big|_{\gamma=\pi/4} = \frac{1}{N!}\, \int_{\IR} \ \prod\limits_{i=1}^{N}\,  \dd\lambda_i \  \e^{t\, \lambda _{i}}\, \frac{\sinh \frac{\lambda _{i}}{4%
}\,\pi }{\sinh \frac{\lambda _{i}}{2}\,\pi } \ \prod_{i<j}\, (\lambda_i-\lambda_j)^2 =\frac{%
G_2(N+1)^{2}}{\big( \sinh ( \frac{%
\pi }{4}+t) \, \sinh ( \frac{\pi }{4}-t) \big)^{N^{2}}}
\end{equation*}%
is essentially the square of the Gaussian matrix model. Notice
that the weight function of this matrix model can also be written as $\omega
\left( \lambda \right) =\e^{t\,\lambda}/2\cosh\frac\lambda4\,\pi$.

In~\cite{Disor} the matrix model (\ref{MP}) is
also studied with the rescaling $\lambda _{i}=N\, \mu _{i}/\gamma $ giving
\begin{equation*}
\tau_{N}=N!\, \Big(\,\frac{ \gamma}{N}\, \Big)^{N^2} \, \int_{\IR} \ \prod\limits_{i=1}^{N} \,  \dd\mu_i \ \e^{-N\, V_N\left( \mu_{i}\right) } \ \prod_{i<j}\, (\mu_i-\mu_j)^2 \ ,
\end{equation*}
where
\begin{equation*}
V_{N}(\mu ) =-\zeta\, \mu -\frac{1}{N}\, \log \Big(\, \frac{\sinh N\, \mu \, \big(
\frac{\pi }{2\gamma }-1\big) }{\sinh N\, \mu\, \frac{\pi }{2\gamma }} \, \Big)
\end{equation*}%
with $\zeta=N\,t/\gamma$. In the thermodynamic limit, $\lim_{N\to\infty}\, V_{N}(\mu )= V(\mu
)=-\zeta \, \mu +\left\vert \mu \right\vert$, this is a linear confining potential, which generally in a one-matrix model lies on the border between
a determined and an undetermined moment problem. This yields a very different behaviour compared to standard matrix models; nevertheless, it is still a
determined moment and hence at least in principle it could be compared to the
Gaussian matrix model. See~\cite{linear} for more details and a way to compute in such
an ensemble with orthogonal polynomials. Finally, let us point out that on the critical line between the ferroelectric and disordered phases the partition function exhibits a novel sub-leading fractional behaviour $N^{1/2}$~\cite{critical}, which is again different from the other phases. Although the precise gauge theory interpretation of this critical line is not yet clear, it is known that certain non-Gaussian matrix models with potentials similar to that of (\ref{MP}) and with a leading $N^{3/2}$ scaling arise from localization of superconformal Chern-Simons theories~\cite{Kapustin:2009kz}.

\subsection{Antiferroelectric phase and Chern-Simons theory on $S^{3}/\mathbb{Z}_{2}$}

Let us finally consider the thermodynamic partition function (\ref{Anti-Z}) in the antiferroelectric region. As mentioned earlier, this type of behaviour
does not correspond to the standard topological expansion of matrix models. It has become familiar more recently in the study of
matrix models with multi-cut domain solutions~\cite%
{Bonnet:2000dz,Eynard:2008yb}. Indeed, the solution in~\cite{Anti} involves
a discrete matrix model; this is the discrete counterpart of the continuous
matrix model that describes the disordered phase, which has a two-cut
solution. Intuitively, the oscillatory behaviour can be related to the
possibility of eigenvalue tunneling from one cut to another \cite{Bonnet:2000dz}; in the case of antiferroelectric order, with pairs of opposing dipoles on the square lattice, this tunneling could have an analog in terms of flipping the polarization of one of the dipoles. The problem of multi-cut solutions of matrix models has
been studied with more detail in \cite{Eynard:2008yb}, where a more general
asymptotic formula is obtained including oscillations to all orders (coming from a
theta-function and its derivatives). As shown in \cite{Eynard:2008he}, and
consistently with rigorous results in the case of the Hermitian one-matrix model,
the oscillatory terms can be resummed in terms of a single theta-function. Various applications of this kind of behaviour in the context of topological string theory are discussed in~\cite{Eynard:2008yb,Eynard:2008he}.

In the context of our gauge theory descriptions of the phases of the six-vertex model with domain wall boundary
conditions in terms of two-dimensional Yang-Mills theory and semiclassical
Chern-Simons theory, the
most precise characterization appears to be in terms of Chern-Simons theory on the lens space $%
L(2,1)=S^{3}/%
\mathbb{Z}
_{2}$. In this case the isomorphism classes of flat $SU(N)$ gauge connections are labelled by $N$-component vectors $(p_1,\dots,p_N)$ with $p_i=0,1$. The partition function is given by
\begin{eqnarray*}
Z_{\rm CS}\big(S^3/\IZ_2\,,\,SU(N) \big)&=& \frac1{(k+N)^{N/2}}\, \sum_{p_i=0,1}\ \prod_{i=1}^N\, \e^{\frac{\pi\ii}2\,(k+i-1)\, p_i} \\ && \qquad \qquad \times \
\prod_{1\leq i<j\leq N}\, 2\sin\frac\pi{2(k+N)}\,\big((k+N)\, (p_i-p_j)+j-i\big) \ .
\end{eqnarray*}
It can be rearranged into a sum over contributions from the $N$ flat connections
$$
Z_{\rm CS}=\sum_{j=0}^{N-1}\, Z_{\rm CS}^{(j)}
$$
which can each be expressed as a matrix integral~\cite{Marino(05)}
\beq
&& Z_{\rm CS}^{(j)}=\frac{\e^{-\frac{\pi\ii}{k+N}\, \rho^2-\frac{\pi^2}{g_s}\,(N-j)}}{j!\, (N-j)!}\, \int_{\IR}\ \prod_{i=1}^N\, \frac{\dd x_i}{2\pi}\ \e^{-\frac1{g_s}\, (x_i-\pi\ii p_i)^2}\ \prod_{i<j}\, \Big(\, 2\sinh\frac{x_i-x_j}2\, \Big)^2 \ ,
\label{ZCSj}\eeq
where $\rho_i=\frac{N+1}2-i$ is the Weyl vector of $SU(N)$, and $(p_1,\dots,p_N)$ is any vector with $j$ entries equal to $0$ and $N-j$ entries equal to $1$. This partition function is related to that of $q$-deformed Yang-Mills theory on $S^2$ given as
\beq
Z_{q{\rm YM}}\big(S^2\,,\,SU(N)\big)= \sum_{n_i\in\IZ}\, \e^{-g_s\, \sum_i\, n_i^2}\ \prod_{i<j}\, \Big(\, 2\sinh\frac{g_s}2\, (n_i-n_j)\, \Big)^2
\label{qYMS2}\eeq
via application of the Poisson resummation formula to express (\ref{qYMS2}) as the instanton expansion~\cite{Caporaso:2005ta,Griguolo:2006kp}
\beq
Z_{q{\rm YM}}=\sum_{j=0}^{N-1}\, \theta_3\big(0\,,\,\e^{-8\pi^2/g_s}\big)^j\, \theta_3\big(\mbox{$\frac{4\pi\ii}{g_s}$}\,,\,\e^{-8\pi^2/g_s}\big)^{N-j}\ Z_{\rm CS}^{(j)} \ ,
\label{ZqYMinst}\eeq
where
$$
\theta _{3}(z,q)=1+2\,\sum_{n=1}^{\infty }\, q^{n^{2}/2}\, \cos \left(n\, z\right) \ .
$$

The contribution (\ref{ZCSj}) to the Chern-Simons partition function is that of a two-cut matrix model. Its large $N$ expansion is studied in~\cite{Marino:2009dp}. In the ``Chern-Simons phase'' wherein $g_s=\frac{2\pi\ii}{k+N}$ is imaginary, the dominant saddle-point contribution to the two-cut free energy for $N$ large and even comes from the ``symmetric'' flat connection labelled by $j=\frac N2$, and to leading orders in this case it is precisely of the form (\ref{Anti-Z}) with~\cite[eq.~(4.18)]{Marino:2009dp}
$$
F_{\rm CS}= \frac{g_s^2}{4N^2}\, F_0\big(\mbox{$\frac{g_s}4\,,\,\frac{g_s}4$}\big)+F_1\big(\mbox{$\frac{g_s}4\,,\,\frac{g_s}4$}\big)+\log\theta_3\big(\,\mbox{$\frac{\pi\ii N}{g_s}\,,\,\e^{F_0^{\prime\prime}}$}\, \big) \ ,
$$
where once again $g_s$ has now been rescaled to the 't~Hooft coupling constant, and the explicit expansions of the genus $0$ and $1$ free energies around $g_s=0$ are given by
\begin{eqnarray*}
F_0(t,t)&=&\frac{\pi^2\,t}2+\log(-4)\, t^2+t^2\,\Big(\log t-\frac34\, \Big)+\frac{t^4}9-\frac{32\,t^6}{575}+\mathcal{O}(t^8) \ , \\[4pt]
F_1(t,t)&=& 2\zeta'(-1)+\frac16\,\log N+\frac{t^2}{72}+\mathcal{O}(t^4) \ .
\end{eqnarray*}
The gauge coupling $g_s$ is related to the weight parameters of the six-vertex model by
$$
F_0^{\prime\prime}=-\frac{\pi^2}\gamma \qquad \mbox{and} \qquad g_s=\frac{\pi\ii\gamma}{2N\,(t+\gamma)+1} \ .
$$
Since this correspondence is made in the 't~Hooft limit, it is again valid in the semiclassical limit $k\to\infty$ of the Chern-Simons gauge theory. Thus while the disordered phase is described by semiclassical Chern-Simons theory on $S^3$, here it is described by the same gauge theory on the three-manifold~$S^3/\IZ_2$.

In the ``Yang-Mills phase'' where $g_s$ is real, the large $N$ expansion for Chern-Simons theory on $S^3/\IZ_2$ is equal to that on $S^3$~\cite[eq.~(4.11)]{Marino:2009dp}, due to exponential suppression of contributions from non-trivial flat connections. It is tempting to relate this behaviour to the spontaneous symmetry breaking from disordered states (with $g_s\in\IR$) to ordered antiferroelectric states (with $g_s\in\ii\IR$) where electric dipoles are alternatingly opposite in each sublattice; the appearance of a local electric field (with overall zero polarization) can then be attributed to the contributions from non-trivial instanton sectors. Note, however, that the proper gauge theory description in this case should be expressed through the $q$-deformed Yang-Mills theory (\ref{qYMS2}), whose instanton expansion (\ref{ZqYMinst}) can still provide the appropriate oscillatory behaviour of the antiferroelectric phase.

The matrix model that describes Chern-Simons theory on the lens space $L(2,1)=S^3/\IZ_2$ is a two-cut matrix model, which is equivalent to $q$-deformed two-dimensional Yang-Mills theory with the discrete matrix model description (\ref{qYMS2}). The matrix model describing the antiferroelectric phase of the six-vertex model is also discrete and is given by the Hankel determinant~\cite{Anti}
$$
\tau_N=\frac{2^{N^2}}{N!}\, \sum_{l_i\in\IZ}\ \prod_{1\leq i<j\leq N}\, (l_i-l_j)^2\ \prod_{i=1}^N\, \e^{2t\, l_i-2\gamma\,|l_i|} \ .
$$
The fact that the Vandermonde determinant here is not the $q$-deformed one of (\ref{qYMS2}) signals that this phase describes the semiclassical $q\to1$ limit. The continuous version of the very same matrix model characterized the disordered phase in \S\ref{Disordered}, and the discussion there surrounding the linear (instead of quadratic) confining potential applies here as well.

\setcounter{section}{0}

\appendix{Painlev\'{e} equations and the Gross-Witten model\label{app:GW}}

The six Painlev\'{e} equations are non-linear differential
equations whose solutions, the Painlev\'{e} transcendents, can be thought of
as nonlinear analogs of the classical special functions~\cite{book1,book2}. The general
solutions of the equations are transcendental since they cannot be
expressed in terms of known elementary functions and, consequently, a new
transcendental function has to be introduced in order to describe their
solutions. The Painlev\'{e} III equation is
\begin{equation}
\frac{\dd^{2}\omega }{\dd z^{2}}=\frac{1}{\omega }\, \Big(\, \frac{\dd\omega }{\dd z}%
\,\Big)^{2}-\frac{1}{z}\, \frac{\dd\omega }{\dd z}+\frac{\alpha\, \omega ^{2}+\beta }{%
z}+\gamma\, \omega ^{3}+\frac{\delta }{\omega }\ ,  \label{1}
\end{equation}%
whereas the Painlev\'e III$^\prime$ equation that appears in e.g.~\cite{ForrWitte} is
\begin{equation}
\frac{\dd^{2}y}{\dd x^{2}}=\frac{1}{y}\, \Big(\, \frac{\dd y}{\dd x}\, \Big) ^{2}-\frac{1}{x}\,
\frac{\dd y}{\dd x}+\frac{\alpha\, y^{2}}{2x^{2}}+\frac{\beta }{2x}+\frac{\gamma
\, y^{3}}{x^{2}}+\frac{\delta }{y} \ . \label{2}
\end{equation}%
Eq.~(\ref{2}) follows from eq.~(\ref{1}) by substituting $\omega \left( z\right) =y(x)/\sqrt{x}$
with $x=\frac{1}{4}\, z^{2}$.

The connection between the Painlev\'{e} equations and the random matrix averages
arising in two-dimensional Yang-Mills theory and the six-vertex model does not directly involve the equations (\ref{1}) or (\ref%
{2}), but rather involves the Hamiltonian formulation of the Painlev\'{e}
equations~\cite{book1,Okamoto}. In this formulation, corresponding to every Painlev\'{e} equation there is a
Hamiltonian $H\left( p,q\right) $ such that, by eliminating the momentum $p$ in the
Hamilton equations, the Painlev\'{e} equation in $q$ follows. One actually has a
family of Hamiltonians $H_{n}$, indexed by a parameter $n$, which in the
correspondence with a random matrix average is identified as the dimension of the matrices. The tau-function is then introduced as a function of an independent variable $t$
and the parameters by%
\[
H_{n}=\frac{\dd}{\dd t}\log \tau _{n} \ .
\]%
Thus the Painlev\'{e} equations themselves are not encountered directly, but rather their $\sigma $-form due to Jimbo, Miwa and Okamoto. In particular, the $\sigma$-form of Painlev\'e~III$^\prime$ reads%
\[
\big( t\, \sigma _{{\rm III}^\prime}^{{\prime \prime }}\big)
^{2}-v_{1}\, v_{2}\, \big( \sigma _{{\rm III}^\prime}^{{\prime }}\big)
^{2}+\sigma _{{\rm III}^\prime}^{{\prime }}\,\big( 4\sigma _{{\rm III}^{\prime
}}^{{\prime }}-1\big)\, \big( \sigma _{{\rm III}^\prime}-t\, \sigma
_{{\rm III}^\prime}^{{\prime }}\big) =\frac{v_{1}-v_{2}}{64} \ ,
\]%
while the $\sigma$-form of Painlev\'e~V is
\beq
&& \big( t\,\sigma _{\rm V}^{{\prime \prime }}\big) ^{2}-\left( \sigma
_{\rm V}-t\,\sigma _{\rm V}^{\prime }+2\left(\sigma _{\rm V}^{\prime }\right) ^{2}+\left(\nu _{0}+\nu _{1}+\nu _{2}+\nu _{3}\right) \,\sigma _{\rm V}^{\prime }\right)
^{2}+4\, \prod\limits_{i=0}^{3}\, \left( \nu _{i}+\sigma _{\rm V}^{\prime }\right) =0 \ .
\label{sigmaPV}\eeq
The Painlev\'{e} equations in $\sigma $-form follow directly from the
Hamiltonian formalism; the Hamiltonian itself satisfies a
differential equation and the $\sigma $-form of the Painlev\'{e} equation is
the equation satisfied by an auxilliary Hamiltonian, defined in terms of the
original Hamiltonian and the parameters. For example, in the case of the
third Painlev\'{e} equation one has
\[
\sigma _{{\rm III}^\prime}^{\prime \prime }(t)=t\, H_{{\rm III}^\prime}+\mbox{$\frac{1}{4}\, \nu _{2}^{2}-\frac{1}{2}\, t$} \ .
\]

The Gross-Witten model is related to the Painlev\'{e} II equation, an important
result in the physics literature~\cite{Periwal:1990gf} that is now known to be a particular case of a more general
mathematical relationship between the matrix model and the theory of Painlev\'{e} transcendents~\cite{Hisakado:1996di,TW}. We have seen in \S\ref{subsec:unitary} that the partition function of the Gross-Witten model is a tau-function of the Painlev\'e~III$^\prime$ equation. In the mathematics literature, the Gross-Witten model (\ref{GW}) is studied as a
Toeplitz determinant $D_{N}(t)=\det_{1\leq i,j\leq N}\,\left[ f_{i-j}\right] $
with symbol $f\left( z\right) =\e^{t\,(z+z^{-1})}$, where $t=\sqrt\theta$.
The usual expression for the partition function in terms of the normalization coefficients $h_n$
of the associated orthogonal polynomials is%
\begin{equation}
D_{N}(t)=\prod\limits_{n=0}^{N-1}\, h_{n} \ . \label{h}
\end{equation}%

If we set $\Phi _{n}=1-U_{n}^{2}=h_{n+1}/h_{n},$ then $\Phi _{n}\left(
t\right) $ satisfies a variant of the Painlev\'{e} V equation
\begin{equation}
\Phi _{n}^{\prime \prime }=\frac{1}{2}\, \Big(\, \frac{1}{\Phi _{n-1}}+\frac{1}{%
\Phi _{n}}\,\Big)\, \left( \Phi _{n}^{\prime }\right) ^{2}-\frac{1}{t}\, \Phi
_{n}^{\prime }-8\Phi _{n}\, \left( \Phi _{n}-1\right) +\frac{2n^{2}}{t^{2}}\, %
\frac{\Phi _{n}-1}{\Phi _{n}} \ . \label{ap5}
\end{equation}
The relationship with the partition function is given by%
\begin{equation*}
D_{N}(t)=\exp \Big(\, 4\int_{0}^{t}\, \dd s~ \log \left( t/s \right)\, s \, \Phi
_{N}\left(s \right) \,\Big) \ .
\end{equation*}%
As shown in \cite{Hisakado:1996di}, eq. (\ref{ap5}) is a Painlev\'{e} V
equation that, using the work of Okamoto \cite%
{Okamoto}, is known to be reducible to Painlev\'{e} III. In \cite{TW} it is also shown that if one
introduces the quantities $W_{n}=U_{n}/U_{n-1},$ then $W_{n}$ satisfies
another special case of the Painlev\'{e} III equation.

This connection to Painlev\'{e} III is related to the well-known result of Periwal
and Shevitz~\cite{Periwal:1990gf} by the consideration of the double scaling
limit $N\rightarrow \infty $ and $t\rightarrow \infty $ with $t/N$ finite, which reduces Painlev\'{e} III to Painlev\'{e} II \cite%
{Hisakado:1996di}. Recall that Painlev\'{e} II actually appears as the
continuum limit of the discrete Painlev\'{e} II equation satisfied by
the coefficients $U_{n}$ as
\begin{equation*}
\left( n+1\right)\, U_{n}=t\, \left( U_{n-1}+U_{n+1}\right)\, \left(
1-U_{n}^{2}\right) \ ,
\end{equation*}%
a property that readily follows from (\ref{h}) together with generic
recursion properties of the normalization coefficients $h_{n}$. This property is also used in
the derivation of the Painlev\'{e} III equation mentioned above \cite{Hisakado:1996di}.

We have seen in this paper that the chiral sector of Yang-Mills theory on $%
\Sigma=S^{2},$ based on the heat kernel action, is intimately related to the
Gross-Witten model which is also a two-dimensional lattice gauge theory,
but on $%
\Sigma=\mathbb{R}
^{2}$ and based on the Wilson action. This relationship only holds for $%
N\rightarrow \infty $. For finite $N$ it is possible to obtain a
relationship between this chiral theory and Painlev\'{e} transcendents by
using mathematical results concerning the Toeplitz determinant (\ref{T}), instead of
that above associated to the Gross-Witten model. Both Toeplitz
determinants are studied in detail in \cite{Bor}.

\end{document}